\documentclass[a4paper,11pt]{article}
\usepackage{jheppub}
\usepackage[T1]{fontenc}
\usepackage{graphicx}
\usepackage{amsmath}
\usepackage{amssymb}
\usepackage{amsfonts}
\usepackage{bm}
\usepackage[table]{xcolor}

\def\be{\begin{equation}}
\def\ee{\end{equation}}
\def\bea{\begin{eqnarray}}
\def\eea{\end{eqnarray}}

\newcommand{\mc}[1]{\mathcal{#1}}
\newcommand{\f}[2]{\frac{#1}{#2}}

\title{\boldmath Cosmology of a higher derivative scalar theory with non-minimal Maxwell coupling}

\author[~]{Shahab Shahidi}

\affiliation[~]{School of Physics, Damghan University, Damghan, 
	41167-36716, Iran}

\emailAdd{s.shahidi@du.ac.ir}

\abstract{
 Higher derivative scalar field theory in curved space-time belongs to the GLPV theory coupled non-minimally to the Maxwell field is considered. We will show that the theory admits two independent exact de Sitter solutions in the FRW background, one driven by the cosmological constant and the other by the GLPV scalar field. The dynamical system analysis of the theory shows that these two exact solutions are stable fixed points. Also, cosmological perturbations over these solutions shows that the cosmological constant based solution is healthy at linear level but the GLPV based solution suffers from a gradient instability in the scalar sector. This proves that the cosmological constant is needed in the GLPV-Maxwell system in order to have a healthy de Sitter solution.
	}
\begin{document}
\maketitle

\section{Introduction}
Modifying Einstein's general relativity has a long history. Perhaps the first modification, can be attributed to the addition of the cosmological constant to the gravitational field equation by Einstein himself \cite{wein}. From then, infinite number of modifications have come out, concentrating on both ultraviolet/infrared limits of the Einstein's field equations \cite{infinite}.  Cosmology however, suffers from many problems, one of the most important is the accelerated expansion of the universe at late times. This can be explained by introducing some light degree of freedom (dof) to the Einstein's field equations, which can be responsible for the IR modification of gravity. Many proposals have been suggested so far in the literature, including the addition of some extra field to the Einstein's theory, which can be a scalar/vector/tensor field \cite{scalar}, or enriching the gravitational action itself like higher order derivative theories \cite{fR}, Weyl-Cartan theories \cite{weyl} or massive gravity theories \cite{massive}.Also one can assume some non-trivial matter-geometry coupling to explain the accelerated expansion of the universe. 

Among all, addition of a scalar field may be the minimal modification of the theory. This adds one additional dof to the Einstein's theory (with two dof) if the Lagrangian for the scalar field is healthy. In order for the scalar interactions to becomes healthy, the scalar field should not have more than two time derivatives at the level of equations of motion, and the interaction terms should have a form which avoid gradient/tachyonic instabilities. The scalar field theories is then divided into two major classes; those which produce accelerated expansion from the kinetic interactions \cite{kesse}, and those which do that from non-trivial potential terms \cite{scalar}.

One the most interesting scalar field theories for the above goal, is the so-called Galileon theory \cite{galileon}. Galileons are scalar fields which has more than second order time derivatives in the action but due to the special form of the interactions, it has at most second order time derivatives in the equations of motion. This makes the theory free from Ostrogradski instability. Galileon terms has an internal symmetry under which the interaction terms remain invariant if one shift the scalars as
$$\phi\rightarrow\phi+b_\mu x^\mu+c,$$
where $\phi$ is the Galileon scalar and $b_\mu$ and $c$ are constants. Many works has been done in the literature, considering cosmological \cite{cosgali}, balck holes \cite{blackgali}, quantum nature \cite{quantgali} and some generalizations of the Galileon scalars \cite{gengali}. However, one of most interesting facts about the Galileons is that they can be interpreted as a position of the 4D brane world embedded in the 5D flat space \cite{branegali}. This suggests that the Galileon interaction terms can not have an arbitrary form and as a result we have a finite number of Galileon interaction terms in any dimension \cite{galileon}.

Upon generalizing the Galileon interactions to curved space time, one immediately find out that higher order time derivatives come back to the equations of motion \cite{covgali}. This is due to the fact that in curved space time, partial derivatives do not commute. This problem can be solved by adding to the action some higher order derivative terms which compensate the higher order time derivatives in the equations of motion. However, these terms breaks the Galileon invariance \cite{covgali}. The most general scalar-tensor interactions in curved space time which has the property that the equations of motion are healthy is called the Horndeski theory \cite{hornde}. Among all the Horndeski terms, four terms bring more attention in the sense that any combination of these terms have a consistent self-tuning mechanism on FRW background. These terms are well-known as the Fab-four \cite{fabfour} an can be written as
\begin{align}
\mc{L}_{john}&=\sqrt{-g}V_{john}(\phi)G^{\mu\nu}\nabla_\mu\phi\nabla_\nu\phi,\qquad\qquad\quad~~ \mc{L}_{george}=\sqrt{-g}V_{george}(\phi)R,
\nonumber\\
\mc{L}_{paul}&=\sqrt{-g}V_{paul}(\phi)P^{\mu\nu\alpha\beta}\nabla_\mu\phi\nabla_\alpha\phi\nabla_\nu\nabla_\beta\phi,\qquad
\mc{L}_{ringo}=\sqrt{-g}V_{ringo}(\phi)\mc{G},
\end{align}
where $P^{\mu\nu\alpha\beta}$ is the double dual of the Riemann tensor and $\mc{G}$ is the Gauss-Bonnet invariant. 

Also it is proposed that the Horndeski theory can be generalized further to contain terms proportional to the Levi-Civita tensor \cite{glpv}
\begin{align}\label{glp}
L_4\supseteq\epsilon_{\mu\gamma\alpha\beta}\epsilon_{\nu\delta\rho}^{~~~~\beta}\nabla^\mu\phi\nabla^\nu\phi\nabla^{\gamma}\nabla^\delta\phi\nabla^\rho\nabla^\alpha\phi,
\end{align}

where $L_4$ is the fourth Horndeski Lagrangian (there is also a similar term for the fifth Horndeski Lagrangian \cite{glpv}). These term will produce third-order derivative terms in the equations of motion but it can be shown that the extra ghost dof does not appear in this case. 
The Fab-four terms can be further generalized in the sense that the potentials for John and Paul terms can depend on $\phi$ and also on $X=\partial_\mu\phi\partial^\mu\phi$. The resulting theory is called beyond Fab-four \cite{beyfab}. This new theory however is a subclass of the GLPV theory, as will be reviewed in the next section.

In this paper, we will investigate cosmological consequences of a scalar field theory coupled to a Maxwell field. The procedure of defining the action is that we write an Einstein-Maxwell system in the presence of the cosmological constant, and then couple the energy momentum tensor of this theory with the kinetic term of the scalar field \cite{babi1}. This will construct the John term of the GLPV theory coupled non-minimally to the Maxwell field. It is well-known that the Fab-four can not satisfy the gravitational wave observations which indicate that the speed of a gravitational wave should be luminal \cite{gw,gw1}. We will then add to the action a term proportional to \eqref{glp} to overcome this difficulty. The result is that the gravitational waves will propagate with the speed of light. We will also see that the theory allow us to have two different exact de Sitter solutions which we will separately investigate the cosmological implications in this paper.
\section{The action}
In this section we will introduce the model and construct the action. This was first done in \cite{babi1}. Let us begin with a gravitational action minimally coupled to the Maxwell field
\begin{align}\label{ac1}
S=\int d^4x\sqrt{-g}\Bigg[\kappa^2 R-2\Lambda-\f14F_{\mu\nu}F^{\mu\nu}\Bigg],
\end{align}
where we have introduced the cosmological constant $\Lambda$ and $F_{\mu\nu}=\partial_\mu A_\nu-\partial_\nu A_\mu$ is the strength tensor related to the electromagnetic potential $A_{\mu}$. Now variation of each term in \eqref{ac1} with respect to the metric tensor gives the Einstein's tensor $G_{\mu\nu}$, the metric tensor $g_{\mu\nu}$ and the energy momentum tensor of the Maxwell field $T_{\mu\nu}$ defined as
\begin{align}
T_{\mu\nu}=\f12F_{\mu\alpha}F_\nu^{~\alpha}-\f18F_{\alpha\beta}F^{\alpha\beta}g_{\mu\nu},
\end{align}
respectively. In this level we can couple a scalar field with the theory \eqref{ac1} by multiplying $f(X)\partial_\mu\phi\partial_\nu\phi$ with the terms obtained from variation of the action \eqref{ac1}. $f$ is an arbitrary function of $X=\partial_\alpha\phi\partial^\alpha\phi$. Note that we allow only the dependence of $f$ on $X$ and not on the field $\phi$ itself, because we want to keep the translational symmetry of the theory, i.e.
$\phi\rightarrow\phi+const.$
The resulting action becomes
\begin{align}\label{ac2}
S=\int d^4x\sqrt{-g}\Bigg[\kappa^2 R&-2\Lambda-\f14F_{\mu\nu}F^{\mu\nu}+f_1(X)\partial_\mu\phi\partial^\mu\phi\nonumber\\&+f_2(X) G^{\mu\nu}\partial_\mu\phi\partial_\nu\phi+f_3(X) T^{\mu\nu}\partial_\mu\phi\partial_\nu\phi\Bigg],
\end{align}
where $f_i$ are arbitrary functions. Let us consider the self interaction term of the scalar field, i.e. the term corresponding to $f_2$. This is a subclass of the beyond Fab four theory \cite{beyfab}, which is known to be a subclass of the GLPV  theory \cite{glpv}. One can see that the term $f_2(X) G^{\mu\nu}\partial_\mu\phi\partial_\nu\phi$ can be reduced to
\begin{align}
f_{2,X}\epsilon_{\mu\gamma\alpha\beta}\epsilon_{\nu\delta\rho}^{~~~~\beta}\nabla^\mu\phi\nabla^\nu\phi\nabla^{\gamma}\nabla^\delta\phi\nabla^\rho\nabla^\alpha\phi+(f_2X)_{,X}\Big[(\Box\phi)^2-\nabla_\mu\nabla_\nu\phi\nabla^\mu\nabla^\nu\phi\Big]-\f12f_2X\,R,
\end{align}
after some integration by parts. This is exactly the fourth beyond Horndeski term \cite{glpv} with identification
$$ G_4=-1/2f_2X,\qquad F_4=-f_{2,X},$$
where $G_4$ and $F_4$ are arbitrary functions introduced in the beyond Horndeski Lagrangian; see \cite{glpv}.

As we have noted in the introduction, recent observational data shows that the speed of the gravitational waves should be equal to the speed of light with an error of order $10^{-15}$ \cite{gw}. In the beyond Horndeski theories, if one considers only the fourth term, one can prove that if the condition
\begin{align}\label{cons}
2G_{4,X}-XF_4=0,
\end{align}
holds, the theory has a tensor mode propagating with the speed of light \cite{gw1}. For the beyond Fab four theory, it is easy to check that the above condition leads to $f_2=0$. As a result the term corresponding to $f_2$ in the action \eqref{ac2}, is not satisfied with the gravitational waves observations. This shows that the scalar field self-interaction term obtained above should be vanishes from the theory. In order to solve this problem, let us add a term 
$$f_{2,X}\epsilon_{\mu\gamma\alpha\beta}\epsilon_{\nu\delta\rho}^{~~~~\beta}\nabla^\mu\phi\nabla^\nu\phi\nabla^{\gamma}\nabla^\delta\phi\nabla^\rho\nabla^\alpha\phi,$$
to the action \eqref{ac2}. The theory then differs from the beyond Fabfour theory but remains a subclass of the GLPV theory. After solving the constraint \eqref{cons} for the new action, one can obtains $f_2=\beta X$, where $\beta$ is an integration constant. Supposing for simplicity that $f_1$ and $f_3$ are constants, one can write the action as
\begin{align}\label{ac21}
S=\int d^4x\sqrt{-g}\Bigg[\kappa^2 R&-2\Lambda-\f14F_{\mu\nu}F^{\mu\nu}+\alpha\partial_\mu\phi\partial^\mu\phi+\beta X G^{\mu\nu}\partial_\mu\phi\partial_\nu\phi\nonumber\\&+\beta\epsilon_{\mu\gamma\alpha\beta}\epsilon_{\nu\delta\rho}^{~~~~\beta}\nabla^\mu\phi\nabla^\nu\phi\nabla^{\gamma}\nabla^\delta\phi\nabla^\rho\nabla^\alpha\phi+\gamma T^{\mu\nu}\partial_\mu\phi\partial_\nu\phi\Bigg],
\end{align}
where $\alpha$ and $\gamma$ are some constants. In the following we will explicitly show that the tensor modes propagate with the speed of light in this theory. One can also rewrite the above action in the form of the GLPV theory as
\begin{align}\label{ac22}
S&=\int d^4x\sqrt{-g}\Bigg[\left(\kappa^2-\f12\beta X^2\right) R-2\Lambda-\f14F_{\mu\nu}F^{\mu\nu}+\alpha\partial_\mu\phi\partial^\mu\phi+\gamma T^{\mu\nu}\partial_\mu\phi\partial_\nu\phi\nonumber\\&+2\beta\epsilon_{\mu\gamma\alpha\beta}\epsilon_{\nu\delta\rho}^{~~~~\beta}\nabla^\mu\phi\nabla^\nu\phi\nabla^{\gamma}\nabla^\delta\phi\nabla^\rho\nabla^\alpha\phi+2\beta X\Big[(\Box\phi)^2-\nabla_\mu\nabla_\nu\phi\nabla^\mu\nabla^\nu\phi\Big]\Bigg],
\end{align}

In order to have a canonical kinetic term for the scalar field, one should set $\alpha=-1/2$. However, we will keep it arbitrary since there is a non-trivial background cosmological solution for $\alpha\neq-1/2$.

As we have discussed above, the theory has a translational symmetry on the scalar field $\phi\rightarrow\phi+a$ with $a$ a constant. Also the above theory has a $U(1)$ symmetry on the Maxwell field $A_{\mu}\rightarrow A_\mu+\partial_\mu\lambda$ with $\lambda$ is an arbitrary function. In this sense, the field equations corresponding to the scalar field $\phi$ and the Maxwell field $A_\mu$ can be written as a conservation of the corresponding Noether charges.
The metric field equation can be written as
\begin{align}\label{met}
G_{\mu\nu}&=T_{\mu\nu}-\Lambda_{eff}\,g_{\mu \nu}-{\beta}\,\bigg(4(\Box\phi)^2\nabla^\alpha\phi\nabla_\alpha\nabla_{(\mu}\phi\nabla_{\nu)}\phi-\f12X^2G_{\mu\nu}+2XR_{\mu\alpha\nu\beta}\nabla^\alpha\phi\nabla^\beta\phi\nonumber\\&+2X(\nabla_\alpha\nabla_\mu\phi\nabla^\alpha\nabla_\nu\phi+\nabla_\alpha\nabla_\mu\nabla_\nu\phi\nabla^\alpha\phi)-4\nabla^\alpha\phi\nabla^\beta\phi\nabla_\alpha\nabla_\beta\nabla_{(\mu}\phi\nabla_{\nu)}\phi\nonumber\\&-8\nabla^\alpha\phi\nabla_\alpha\nabla_\beta\phi\nabla^\beta\nabla_{(\mu}\phi\nabla_{\nu)}\phi\bigg)- 2\Big(\alpha-(\Box\phi)^2+2(\nabla\nabla\phi)^2+\nabla^\alpha\phi\Box\nabla_\alpha\phi\nonumber\\&+G_{\alpha\beta}\nabla^\alpha\phi\nabla^\beta\phi\Big)\nabla_{\mu}\phi\nabla_{\nu}\phi-\frac{1}{2}\gamma\bigg(F_{\mu\alpha}F_{\nu \beta}
\nabla^{\alpha }\phi\,\nabla^{\beta }\phi+2\nabla_{(\mu}\phi F_{\nu) \alpha}
F^{\beta \alpha} \nabla_{\beta}\phi\nonumber\\&-\frac{1}{2}\,(\nabla\phi)^2F_{\mu \sigma} F_{\nu}^{\phantom{\nu}
	\sigma}\,
-\frac{1}{4}\,\nabla_{\mu}\phi\,\nabla_{\nu}\phi\,F^2\bigg),
\end{align}
where we have defined
\begin{align}
\Lambda_{eff}&=\Lambda-\alpha X-2\beta\Big[X\nabla^\alpha\phi\Box\nabla_\alpha\phi+X(\nabla\nabla\phi)^2-\nabla^\alpha\phi\nabla^\beta\phi\nabla^\gamma\phi\nabla_\gamma\nabla_\alpha\nabla_\beta\phi\nonumber\\&-\nabla^\alpha\phi\nabla^\beta\phi\nabla_\gamma\nabla_\alpha\phi\nabla^\gamma\nabla_\beta\phi\Big]+\f14\gamma\Bigg[\frac{1}{4}XF^2-F_{\beta \alpha}F_{\tau}^{ \phantom{\tau} \alpha}\,\nabla^{\beta }\phi\,\nabla^{\tau }\phi\Bigg],
\end{align}
and we have used the following notation
\begin{align}
 (\nabla\nabla\phi)^2\equiv\nabla_\mu\nabla_\nu\phi\nabla^\mu\nabla^\nu\phi,\quad F^2\equiv F_{\mu\nu}F^{\mu\nu}.
\end{align}
The scalar and vector field equations can be written respectively as
\begin{align}\label{scal}
\nabla_{\mu}\bigg[  \big(\alpha\,g^{\mu\nu}&+\beta \,
G^{\mu\nu}X+\beta G_{\alpha\beta}\nabla^\alpha\phi\nabla^\beta\phi g_{\mu\nu}+\gamma\,T^{\mu \nu}\big)
\nabla_{\nu}\phi\nonumber\\&+2\beta \epsilon_{\mu\gamma\alpha\beta}\epsilon_{\nu\delta\rho}^{~~~~\beta}\nabla^\rho\nabla^\alpha\phi\nabla^\gamma\nabla^{[\delta}\phi\nabla^{\nu]}\phi\bigg]  =0,
\end{align}
and
\begin{align}\label{vect}
\nabla_{\mu}\left[\left(1+\f12\gamma(\nabla\phi)^2\right)F^{\mu\nu}+2\gamma F_{\sigma}^{(\mu}\nabla^{
	\nu)}\phi\nabla^{\sigma}\phi\right]=0.
\end{align}
As we have discussed the last two equations of motion can be written in the form $\partial_\mu(\sqrt{-g}J^\mu)=0$ which is the the conservation equations related to translational and $U(1)$ symmetries.
\section{Background cosmology}\label{back}
Let us now consider the cosmological consequences of the model \eqref{met}-\eqref{vect}. Let us assume that the universe can be described by the FRW ansatz with line element
\begin{align}
ds^2=-dt^2+a^2(dx^2+dy^2+dz^2),
\end{align}
where $a=a(t)$ is the scalar factor. In the case of isotropic and homogeneous space-time, the vector field  $A_\mu$ should have the form
\begin{align}
A_\mu=(A_0(t),0,0,0),
\end{align}
and the scalar field can be written as $\phi=\phi(t)$. The field equations then reduces to
\begin{align}\label{eq1}
-3 \kappa ^2 H ^2+\Lambda
-\frac{1}{2} \alpha  \dot{\phi} ^2+\f{15}{2}\beta H^2\dot{\phi}^4=0,
\end{align}
\begin{align}\label{eq2}
-\kappa ^2 (2\dot{H}&+3
H ^2)+\Lambda +\beta  \dot{H}  \dot{\phi} ^4+\frac{3}{2} \beta 
H ^2 \dot{\phi} ^4+4 \beta  H  \dot{\phi}^3   \ddot{\phi}  +\frac{1}{2} \alpha  \dot{\phi} ^2=0,
\end{align}
\begin{align}\label{eq3}
2 \alpha   \ddot{\phi}&+6 \alpha 
H  \dot{\phi} -36 \beta  H ^2  \ddot{\phi}\dot{\phi}^2 -12 \beta  H ^3
\dot{\phi}^3 -24 \beta  H  \left(\dot{H} +H ^2\right) \dot{\phi}^3=0,
\end{align}
where $H$ is the Hubble parameter. Note that the vector field equation of motion is satisfied identically in the case of homogeneous and isotropic universe since our theory is $U(1)$ invariant. Also note that in the above field equations,  the scalar field appears at most with two time derivatives, denoting that the theory does not have an Ostrogradski instability.

The above set of equations has two exact solutions corresponding to an accelerated expanding universe. The first one has non-vanishing cosmological constant, with
\begin{align}\label{lambdabased}
\phi=\phi_0,\qquad H=\sqrt{\f{\Lambda}{3\kappa^2}},
\end{align}
where $\phi_0$ is an arbitrary constant. We refer to this solution as $\Lambda$-based solution. This solution is nothing but the standard dS solution of the Einstein-Hilbert theory with non-vanishing cosmological constant. This happens actually because we have assumed that the scalar field is constant and the field equation contains at least first order time derivatives of the scalar field. So, the scalar field will be disappeared from the equations. Despite the fact that the background solution is the same as in the standard Einstein's theory we will see that at the level of perturbations the physics becomes different from that of Einstein's theory.

The theory has another dS solution with vanishing cosmological constant $\Lambda=0$ and
\begin{align}\label{galileonbased}
\phi=\left(\f{2\kappa^2}{3\beta}\right)^{1/4}\,t,\qquad H=\left(\f{\alpha^2}{24\beta\kappa^2}\right)^{1/4},
\end{align}  
which we will refer as the GLPV-based solution. In this case $\alpha$ and $\beta$ should be positive constants.
This is actually the non-trivial solution of the GLPV-Maxwell system and the accelerated expansion comes from the scalar field. Note that the Maxwell field does not contribute to the background solutions since as noted above, we have assumed isotropic and homogeneous universe. In order to investigate the effects of the Maxwell field, one should consider for example anisotropic space-times.
\subsection{Dynamical system analysis}
\begin{figure}\label{fig4}
	\centering
	\includegraphics[scale=0.5]{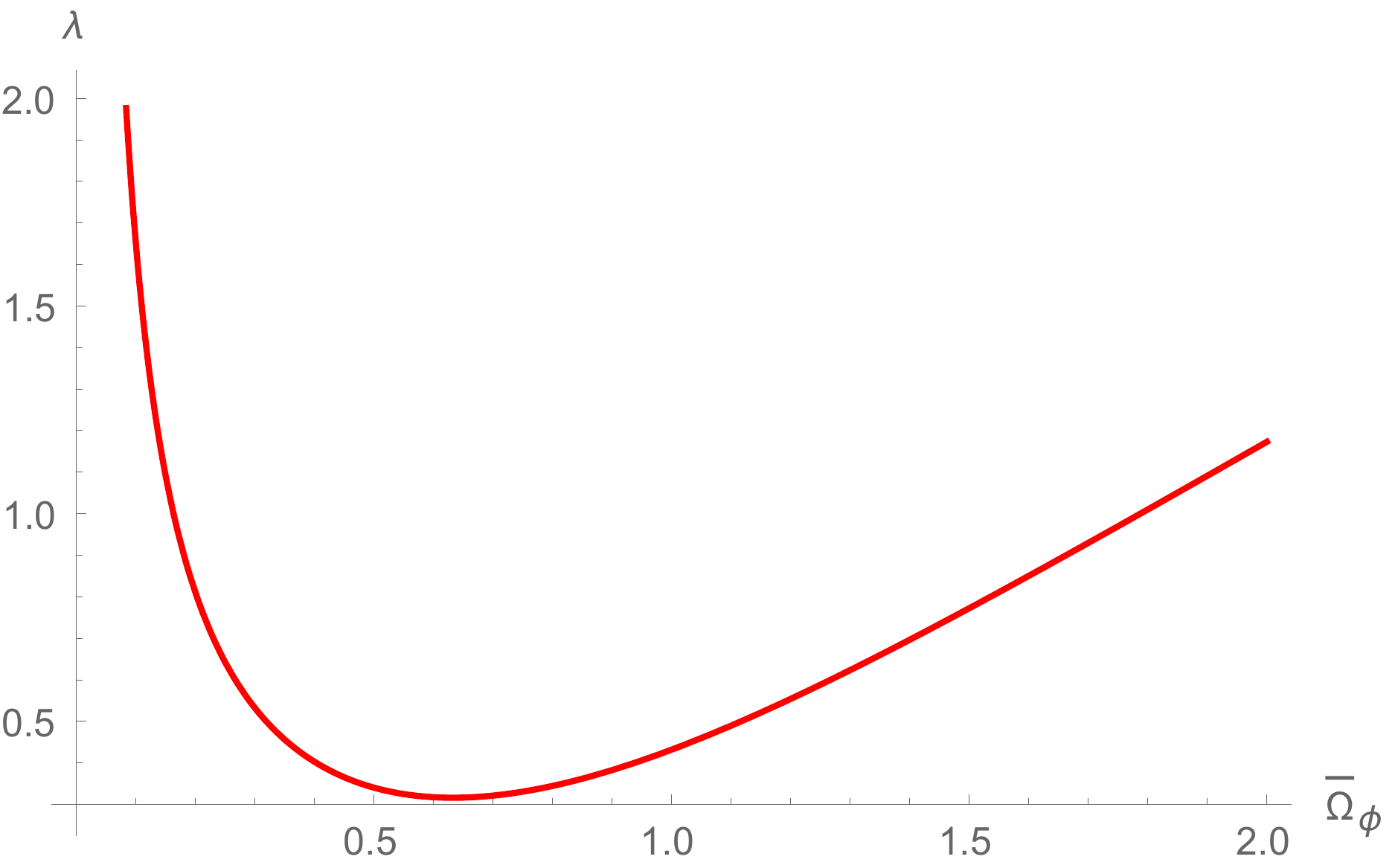}
	\caption{Singular points of the dynamical evolution of the theory. The plot corresponds to the values of $\tilde{\Omega}$ and $\tilde{\lambda}$ which leads to $\mc{A}=0$.}
\end{figure}

Let us write the Friedman equation \eqref{eq1} as
\begin{align}\label{eqq1}
-\frac{\alpha  \dot{\phi}^2}{6 \kappa ^2 H^2}+\frac{\kappa ^2 \bar{\Lambda}}{3
	H^2}+\frac{5\bar{\beta} \dot{\phi}^4}{2 \kappa ^8}=1,
\end{align}
where we have defined dimensionless constants $\bar{\beta}=\kappa^6\beta$ and $\Lambda=\kappa^4\bar{\Lambda}$. From the above equation one can define two dynamical variables as
$$\Omega_\Lambda=\f{\kappa^2}{3H^2},\qquad \Omega_\phi=\f{\dot{\phi}^2}{\kappa^4}.$$
Equation \eqref{eqq1} shows that $\Omega_\Lambda$ can be obtained as a function of $\Omega_\phi$ so the system has only one dynamical degree of freedom. Using equations \eqref{eq2} and \eqref{eq3} one can write the autonomous equation of this degree of freedom as
\begin{align}\label{dyna}
\f{d\Omega_\phi}{d\ln a}=-\frac{6 \Omega _{\phi } \left(5 \bar{\beta } \Omega _{\phi }^2-2\right) \left(\alpha 
	\left(3 \bar{\beta } \Omega _{\phi }^2-2\right)+4 \bar{\beta } \bar{\Lambda } \Omega
	_{\phi }\right)}{\alpha  \left(15 \bar{\beta }^2 \Omega _{\phi }^4+4\right)-4
	\bar{\beta } \bar{\Lambda } \Omega _{\phi } \left(5 \bar{\beta } \Omega _{\phi
	}^2+6\right)}.
\end{align}
One should impose that the denominator of the above expression is non-zero in order to have a non singular cosmological evolution. In figure \eqref{fig4}, we have plotted the singular points which should be excluded from the theory. We have defined $\tilde{\Omega}_\phi=\sqrt{\bar{\beta}}\Omega_\phi$ and $\lambda=\sqrt{\bar{\beta}}\bar{\Lambda}/\alpha$, so that the only parameter to be discussed is $\lambda$. The figure then shows the set of values $(\lambda,\tilde{\Omega}_\phi)$ which leads to the vanishing denominator.

The effective equation of state parameter $\omega_{eff}=-1-2\dot{H}/3H^2$ can be obtained as
\begin{align}
\omega_{eff}=-1-\frac{2 \Omega _{\phi } \left(\alpha  \left(3 \bar{\beta } \Omega _{\phi }^2-2\right)+4
	\bar{\beta } \bar{\Lambda } \Omega _{\phi }\right) \left(\alpha  \left(5 \bar{\beta }
	\Omega _{\phi }^2+2\right)-20 \bar{\beta } \bar{\Lambda } \Omega _{\phi
	}\right)}{\left(\alpha  \Omega _{\phi }-2 \bar{\Lambda }\right) \left(\alpha 
	\left(15 \bar{\beta }^2 \Omega _{\phi }^4+4\right)-4 \bar{\beta } \bar{\Lambda }
	\Omega _{\phi } \left(5 \bar{\beta } \Omega _{\phi }^2+6\right)\right)}
\end{align}
with

In order to have a non singular effective equation of state parameter, one should exclude the points in figure \eqref{fig4} and also a point $\Omega_\phi=2\bar{\Lambda}/\alpha$.

The above system has five fixed points as we will discuss in the following.
\subsubsection{de Sitter fixed points}
The theory \eqref{dyna} has three fixed points corresponding to the de Sitter expansion. The first one is
\begin{align}
\Omega_\phi=0,\quad\Omega_\Lambda=\f{1}{\bar{\Lambda}},\qquad \omega_{eff}=-1,
\end{align}
which is exactly the $\Lambda$-based de Sitter solution obtained in the previous section. The Eigenvalue corresponding to this fixed point is $-6$ indicating that the fixed point is stable.

The second fixed point is
\begin{align}
\Omega_\phi=-\f{2\alpha}{2\bar{\beta}\bar{\Lambda}+\sqrt{2\bar{\beta}(3\alpha^2+2\bar{\beta}\bar{\Lambda}^2)}},\qquad \omega_{eff}=-1,
\end{align}
with Eigenvalue $-3$ indicating that it is stable. Also note that the expression under the square root is always positive. In the limit $\Lambda\rightarrow0$, one obtains $\Omega_\Lambda=(8\bar{\beta}/3\alpha^2)^{1/2}$, which is exactly the behavior of the GLPV-based de Sitter solution of the previous section. As a result both de Sitter solutions obtained analytically in the previous section are dynamically stable. However, there is another stable de Sitter fixed point corresponding to
\begin{align}
\Omega_\phi=-\f{1}{3\alpha\beta}\left(2\bar{\beta}\bar{\Lambda}+\sqrt{2\bar{\beta}(3\alpha^2+2\bar{\beta}\bar{\Lambda}^2)}\right),\qquad \omega_{eff}=-1,
\end{align}
with 
$$\Omega_\Lambda=-\f{2}{3\alpha^2}\left(2\bar{\beta}\bar{\Lambda}+\sqrt{2\bar{\beta}(3\alpha^2+2\bar{\beta}\bar{\Lambda}^2)}\right)$$
and Eigenvalue $-3$. One can see that the above fixed point leads to imaginary Hubble parameter and therefore is not well-defined.
\subsubsection{Matter dominated fixed points}
The theory \eqref{dyna} has also two unstable fixed points correspond to the matter dominated universe. For both fixed points, one has $\Omega_\Lambda=0$, $\omega_{eff}=0$ and the Eigenvalues are $+3$. However, these fixed points differs from the value of $$\Omega_\phi=\pm\sqrt{\f{2}{5\bar{\beta}}}$$
In summary, one has a vast cosmological dynamics in this model. Starting from either unstable matter dominated fixed points and end in one of the stable de Sitter fixed points corresponding to the $\Lambda$-based or GLPV-based solution.
\subsection{General solutions}
\begin{figure}\label{fig1}
	\includegraphics[scale=0.4]{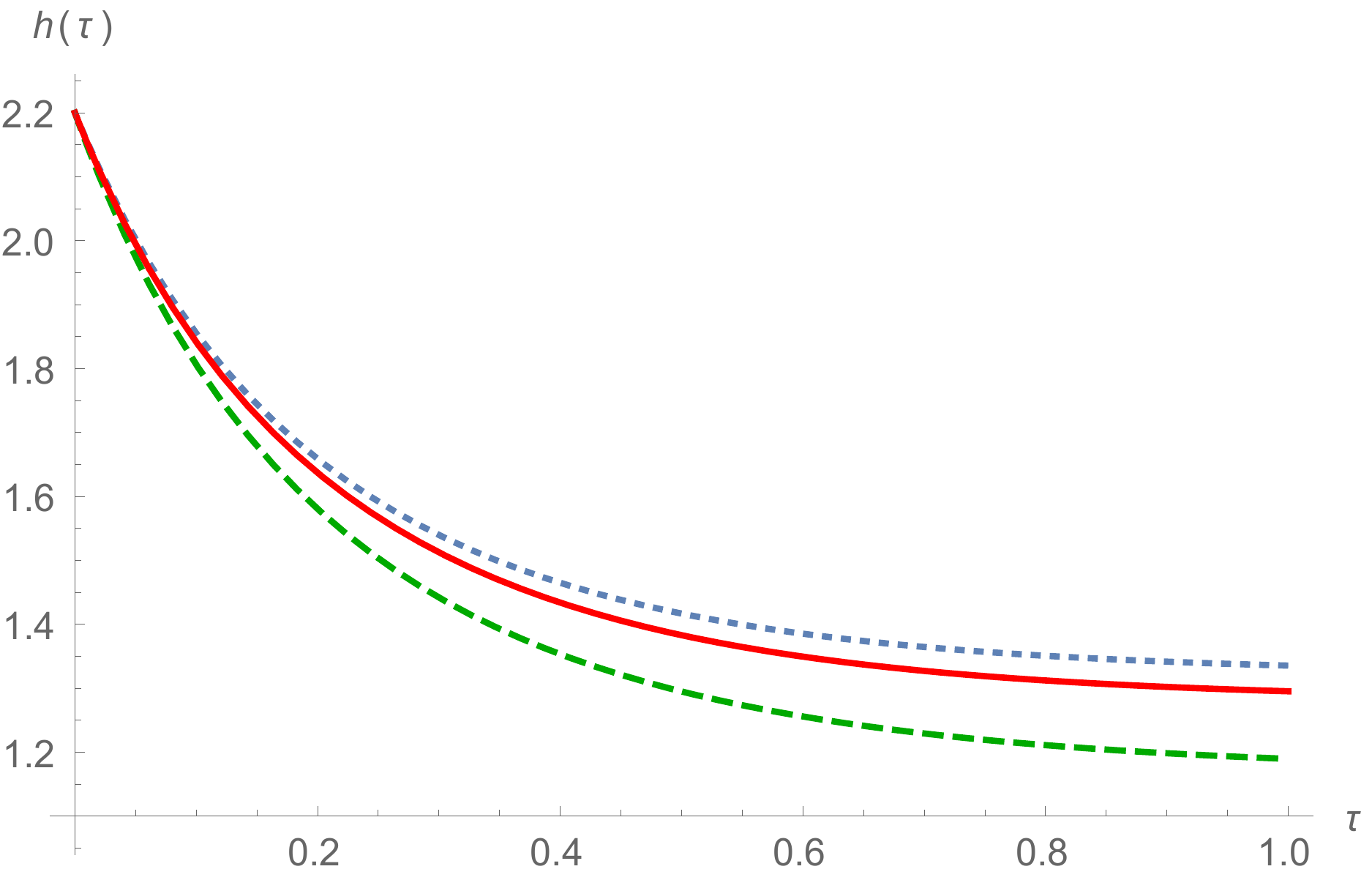}\includegraphics[scale=0.4]{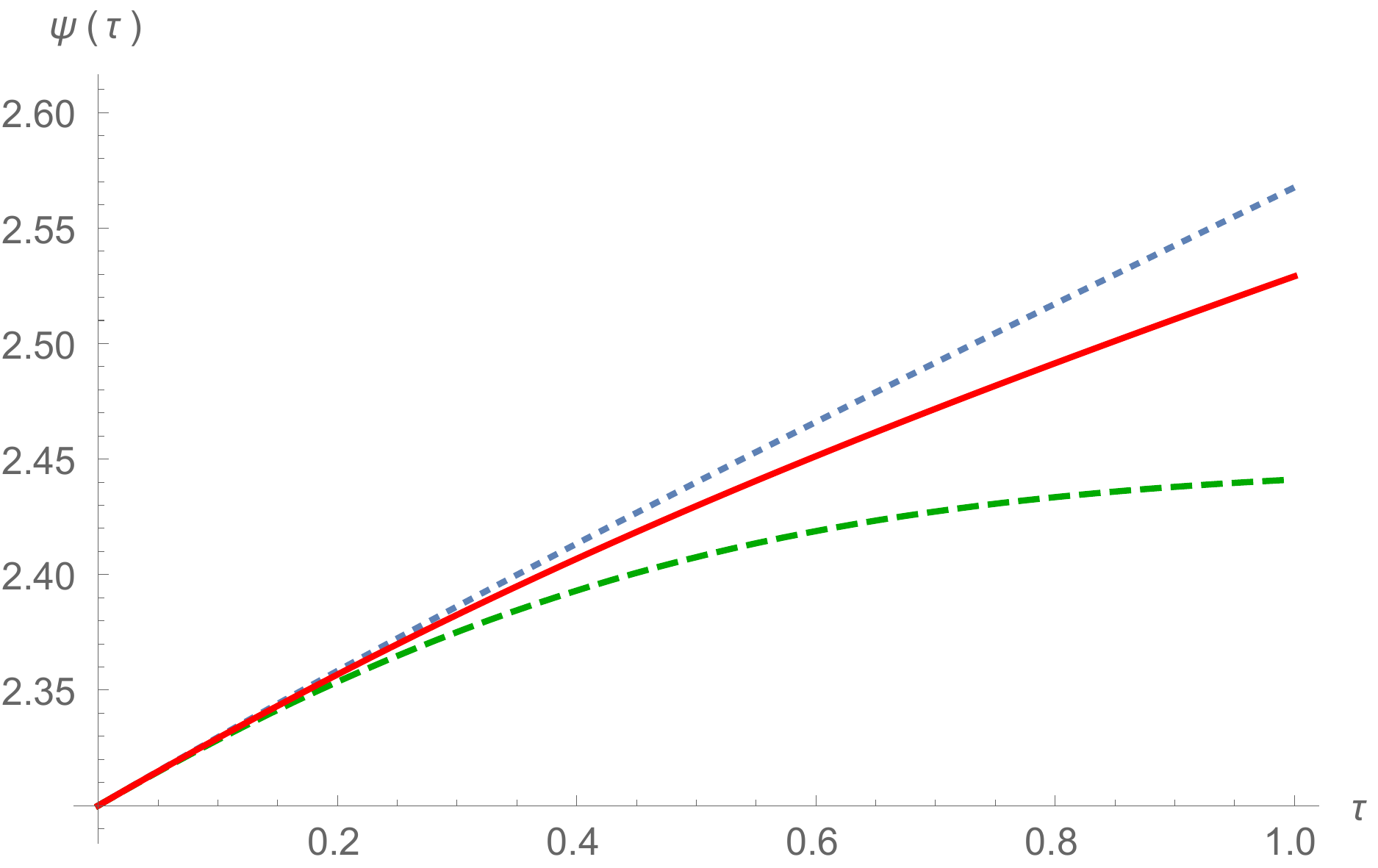}
	\vspace{0.1cm}\\
	\centering\includegraphics[scale=0.43]{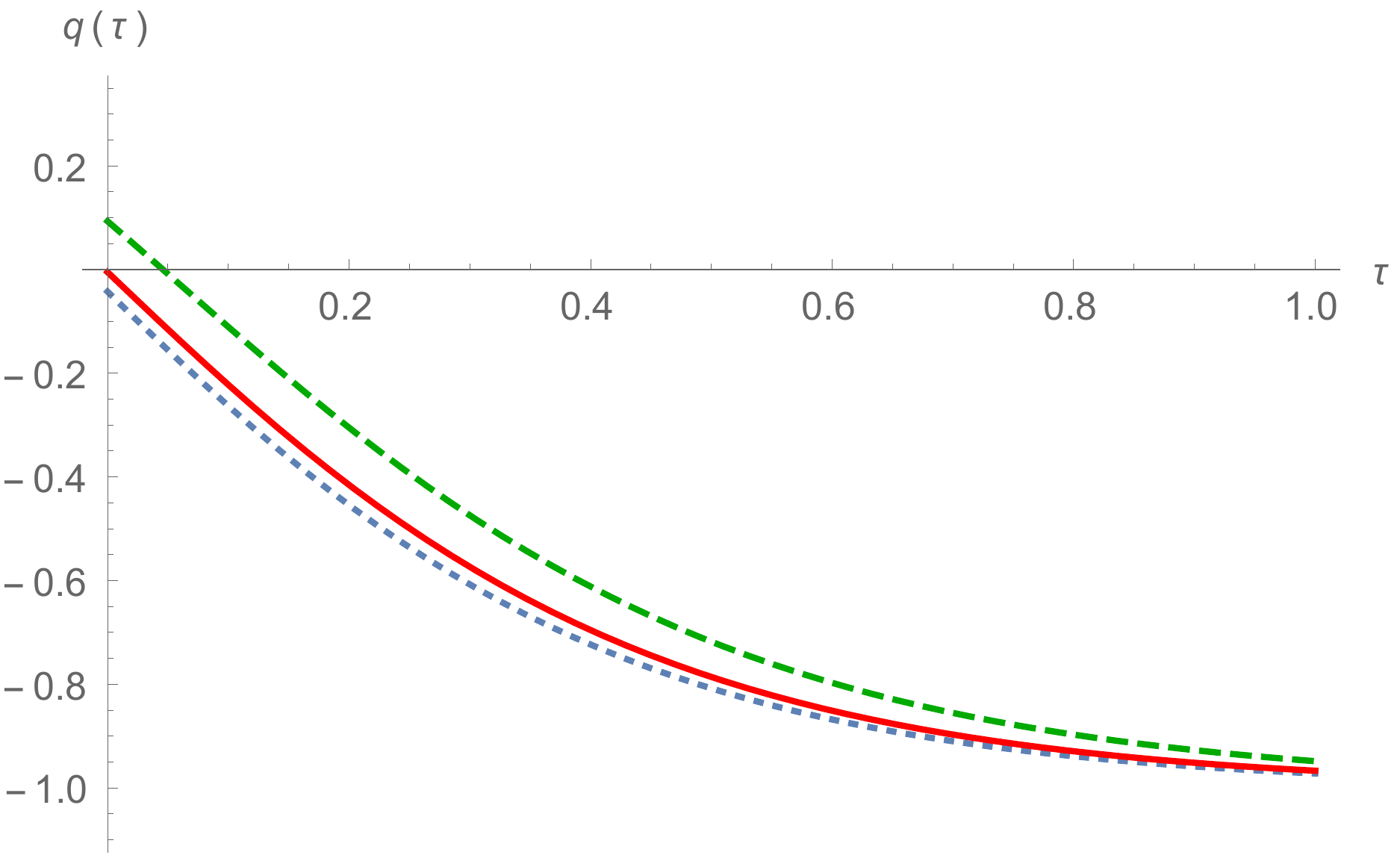}
	\caption{Plot of the Hubble parameter, the scalar field and the deceleration parameter as a function of the dimensionless time parameter $\tau$. The values of the parameters are $\alpha=-1.1,1.1,1$, $\bar{\beta}=2.1,1.7,3.4$ and $\lambda=4.1,5.2,4.9$ for the dotted/dashed/solid lines respectively.}
\end{figure}

Before considering the cosmological perturbations of the above exact solutions, let us solve the system \eqref{eq1}-\eqref{eq3} numerically. Defining the dimensionless parameters
\begin{align}
H=H_0 h,\quad \tau=H_0 t,\quad \phi=\kappa \psi,\quad \Lambda=\lambda \kappa^2 H_0^2, \quad \bar{\beta}=H_0^4\kappa^2\beta,
\end{align}
one can rewrite the equations of motion as
\begin{align}
&h ^2 \left(15 \bar{\beta} \psi^{\prime 4}-6\right)-\alpha \psi^{\prime 2}+2 \lambda=0,\nonumber\\
&3 h  \psi '  \left(4\bar{\beta}\psi^{\prime2}h' -\alpha\right)+18 \bar{\beta}\psi^{\prime2} h ^2 \psi '' +18\bar{\beta}\psi^{\prime3} h ^3 -\alpha \psi'' =0,\nonumber\\
&(2 h'+3 h ^2)  \left(\bar{\beta} \psi^{\prime4}-2\right)+\alpha \psi^{\prime2}+2 \lambda+8\bar{\beta}h\psi^{\prime3}\psi''=0,
\end{align}
where prime denotes derivative with respect to $\tau$. Figure \eqref{fig1} shows the behavior of the Hubble parameter, the scalar field and the deceleration parameter defined as $q=-1-\dot{H}/H^2$ as a function of $\tau$. 
The value of the parameter $\alpha$ is $\alpha=-1.1,1.1,1$ for the dotted, dashed and solid lines respectively. One can see from the figures that both positive and negative values of the parameter $\alpha$ can result in the accelerated expanding universe.
\section{Perturbations}
In this section we will investigate the cosmological perturbations around the background solutions introduced in section \ref{back}. 
The metric perturbations around FRW background can be written as
\begin{align}\label{100}
ds^2=-(1+2\varphi)\,dt^2+2a(S_i+\partial_i B)dx^i\, dt+a^2\big((1+2\psi)\delta_{ij}+\partial_i\partial_j E+\partial_{(i}F_{j)}+h_{ij}%
\big)dx^i dx^j,
\end{align}
where $\varphi$, $\psi$, $E$ and $B$ are the scalar perturbations, $S_i$ and $F_i$ are the vector perturbations with vanishing divergence $\partial_i S_i=0=\partial_i F_i$, and $h_{ij}$ is the traceless and transverse tensor perturbation,
$h_{ii}=0=\partial_i h_{ij}$. Note that in our notation the spatial indices are raised and lowered by the flat-space metric $\delta_{ij}$. 
The Maxwell field can be decomposed as
\begin{align}\label{101}
A_\mu=(A_0+\delta A_0,\xi_i+\partial_i \delta A),
\end{align}
where $A_0$ is the background value of the Maxwell field. Note that due to $U(1)$ symmetry of the action \eqref{ac2}, the Maxwell field did not appear in the background field equations and $A_0$ remains an arbitrary function. In this section for simplicity we will assume that $A_0$ is a constant. In the decomposition of the Maxwell field \eqref{101},  $\delta Q_0$ and $\delta Q$ are the scalar perturbations and $\xi_i$ is a transverse vector perturbation $\partial_i\xi_i=0$.
The scalar field can also be decomposed as
\begin{align}\label{102}
\phi=\phi_0+\delta\phi.
\end{align}
Note that $\phi_0$ is not constant in the GLPV-based solution. 

Now, let us define the gauge invariant perturbation quantities. Under the infinitesimal coordinate transformation of the form 
$x^\mu\rightarrow x^\mu+\delta x^\mu$, the scalar perturbations transform as
	\begin{align}
	&\varphi\rightarrow\varphi-\partial_t\delta x^0,\qquad\quad
	B\rightarrow B+\f{1}{a}\delta x^0-a\partial_t\delta x,\qquad
	\psi\rightarrow\psi-H\delta x^0,\qquad
	E\rightarrow E-2\delta x,\nonumber\\
	&\delta A\rightarrow \delta A-A_0\delta x^0,\quad~
	\delta A_0\rightarrow \delta A_0-A_0\partial_t\delta x^0,\qquad~
	~\delta\phi\rightarrow\delta\phi-\dot{\phi_0}\delta x^0.
	\end{align}
We can construct five gauge invariant scalar perturbations as
\begin{align}
\Phi&=\varphi+\partial_t\left(aB-\f{a^2}{2}\partial_t E\right),\qquad\qquad
\Psi=\psi+H\left(aB-\f{a^2}{2}\partial_t E\right),\nonumber\\
\mathcal{X}&=\delta A_0+A_0\partial_t\left(aB-\f{a^2}{2}\partial_t E\right),\qquad
\mathcal{Y}=\delta A+A_0\left(aB-\f{a^2}{2}\partial_t E\right),\nonumber\\
\mathcal{Z}&=\delta\phi+\dot{\phi_0}\left(aB-\f{a^2}{2}\partial_t E\right).
\end{align}
Note that for the $\Lambda$-based solution the scalar perturbation $\delta\phi$ is gauge invariant and we have $\mathcal{Z}=\delta\phi$.

For the vector perturbation we have
\begin{align}
	S_i\rightarrow S_i-a\partial_t\eta_i,\qquad
	F_i\rightarrow F_i-2\eta_i,\qquad
	\xi_i\rightarrow\xi_i,
	\end{align}
	and we can construct two gauge invariant vector perturbations of the form
	\begin{align}
	\rho_i=S_i-\f12a\partial_t F_i,\qquad \xi_i\rightarrow\xi_i.
	\end{align}
	The tensor perturbation $h_{ij}$ does not transform under the infinitesimal coordinate transformation and so it is gauge invariant.
	\subsection{Tensor perturbation}
	Let us consider the tensor perturbation of the theory \eqref{ac2}. The tensor perturbation $h_{ij}$ has two polarizations which we will denote by $h_\times$ and $h_+$. After expanding the action up to second order in $h_{ij}$ and Fourier transforming the resulting action one obtains
	\begin{align}
	S^{(2)}_{tensor}=\f12\sum_{+,\times}\int\, d^3k\,dt\, \kappa^2\,a^3 a_1\left[\dot{h}_{ij}\dot{h}_{ij}-\f{\vec{k}^2}{a^2}h_{ij}h_{ij}\right],
	\end{align}
	where $a_1=1$ for $\Lambda$-based solution and 
	$a_1=1/6$ for GLPV-based solution. One can see from the above expression that the speed of the tensor modes in both solutions is equal to the speed of light, in agreement with recent gravitational wave observation \cite{gw}.
	
	One should note that the scalar and vector interaction terms does not contribute to the tensor perturbation in the $\Lambda$-based solution since the background values $\phi_0$ and $A_0$ are constant. So, tensor modes in $\Lambda$-based solution is equivalent to the Einstein's theory. However, for the GLPV-based solution where the background value of the scalar field depends on time one has a tensor contribution from the $\beta$ terms in the action \eqref{ac2}.
	
\subsection{Vector perturbation}
For the vector perturbation we have two gauge invariant quantities. After Fourier transformation, one can obtain the vector part of the second order perturbed action as
\begin{align}\label{vecpert}
S^{(2)}_{vector}=&\f12\int d^3kdt a\Bigg[a_1\dot{\xi}_i^2-a_2\f{\vec{k}^2}{a^2}\xi_i^2+a_3\kappa^2\vec{k}^2\rho_i^2\Bigg],
\end{align}	
where $a_1=a_2=a_3=1$ for $\Lambda$-based solution and $$a_1=1+\f{\gamma\kappa}{\sqrt{6\beta}},\quad a_2=1-\f{\gamma\kappa}{\sqrt{6\beta}},\quad a_3=\f{2}{3}$$
 for GLPV-based solution. Note that $\rho_i$ is non-dynamical with equation of motion $\rho_i=0$, so the third term in \eqref{vecpert} vanishes and one obtains the vector perturbation action as
\begin{align}
S^{(2)}_{vector}=&\f12\int d^3kdt\bigg[a_1\dot{\xi}_i^2-a_2\f{\vec{k}^2}{a^2}\xi_i^2\bigg].
\end{align}	
One can see from the above relation that the $\Lambda$-based solution is always healthy. For the GLPV-based solution, noting that $\beta>0$ from \eqref{galileonbased}, the stable vector perturbation implies $\beta>\gamma^2\kappa^2/6$.

\subsection{Scalar perturbation}
For the scalar perturbation, there are five gauge invariant scalar quantities. In what follows we will consider the scalar perturbations over two background solutions separately.
\subsubsection{$\Lambda$-based solution}
After Fourier transformation of the second order action, one obtains
\begin{align}\label{scal1}
S^{(2)}_{scalar}&=\f12\int\, d^3k\,dt\,\Bigg[
\Phi  \left(8 \sqrt{3\Lambda} \mathit{a}^3 \kappa \dot\Psi +8 \mathit{a} \kappa ^2 k^2 \Psi
\right)-2k^2
\mathit{a}\mathcal{X} \dot{\mathcal{Y}}-4\Lambda  \mathit{a}^3 \Phi ^2+4
\mathit{a} \kappa ^2 k^2 \Psi ^2\nonumber\\&+\mathit{a} \left(k^2\dot{\mathcal{Y}{}}^2-2\alpha \mathit{a}^2 \dot{\mathcal{Z}{}}^2-12 \kappa ^2 \mathit{a}^2\dot\Psi {}^2\right)+ 2 \alpha  k^2\mathit{a}
\mathcal{Z}^2+\mathit{a}  k^2\mathcal{X}^2
\Bigg].
\end{align}
One can see from the above action that $\Phi$ and $\mathcal{X}$ are non-dynamical with equations of motion
\begin{align}\label{sol1}
\mathcal{X}=\dot{\mathcal{Y}},\qquad \Phi=\sqrt{\f{3\kappa^2}{\Lambda}}\dot{\Psi}+\f{\kappa^2 k^2}{\Lambda a^2}\Psi.
\end{align}
Substituting back the solutions \eqref{sol1} to the action \eqref{scal1} one can see that $\mathcal{Y}$ vanishes from the action and also $\Psi$ becomes non-dynamical with equation of motion $\Psi=0$. At the end we have left with an action with one scalar dynamical degree of freedom
\begin{align}
S^{(2)}_{scalar}&=-\f12\alpha\int\, d^3k\,dt\,a^3\Bigg[\dot{\mathcal{Z}}^2-\f{\vec{k}^2}{a^2}\mathcal{Z}^2\Bigg].
\end{align}
In order to have a healthy scalar perturbation on top of the $\Lambda$-based solution one should have $\alpha<0$. This in fact expectable since in the $\Lambda$-based solution, only the $\alpha$ term contribute and the kinetic term of the scalar field becomes positive only for $\alpha<0$.
\subsubsection{GLPV-based solution}
In this subsection, we will concentrate on the scalar perturbation over the de Sitter background of the GLPV-based solution \eqref{galileonbased}. After Fourier transforming the second order action, one obtains 
\begin{align}\label{scal2}
S^{(2)}_{scalar}&=\f12\int\, d^3k\,dt\,\frac{1}{6 \mathit{a}^2 \sqrt{\beta }}\Bigg[\sqrt{6} \kappa  \left(\gamma  k^2\dot{\mathcal{Y}} \left(\dot{\mathcal{Y}}-2
	\mathcal{X}\right)+20 \mathit{a}^2 \alpha  \Phi ^2+\gamma  k^2 \mathcal{X}^2\right)\nonumber\\&+2 \sqrt{\beta } \Big(3 k^2 \dot{\mathcal{Y}} \left(\dot{\mathcal{Y}}-2 \mathcal{X}\right)+12 \mathit{a}^2 \left(\alpha \dot{\mathcal{Z}}{}^2-2 \kappa ^2 \dot\Psi {}^2+4 \sqrt{\alpha } \kappa  \dot\Psi \dot{\mathcal{Z}}\right)\nonumber\\&+k^2 \left(3 \mathcal{X}^2+8 \left(\kappa ^2 \Psi  (6 \Phi +\Psi )-\alpha  \mathcal{Z}^2+2
	\sqrt{\alpha } \kappa  \mathcal{Z} (\Psi -2 \Phi )\right)\right)\Big)\nonumber\\&+32(24\beta^3\kappa^6)^{1/4}
	k^2 \mathcal{Z} \dot\Psi -16 (54\beta\alpha^2\kappa^2)^{1/4} \mathit{a}^2 \Phi
	\left(\kappa  \dot\Psi +2 \sqrt{\alpha }\dot{\mathcal{Z}}\right)\Bigg].
\end{align}
It is evident that $\Phi$ and $\mc{X}$ are non-dynamical variables with equations of motion
\begin{align}
\Phi=\f{2}{15}\sqrt{\f{6\beta}{\alpha^2}}(2\sqrt{\alpha}\mc{Z}-3\kappa\Psi)\f{k^2}{a^2}+\f15\left(\f{6\beta}{\alpha^2\kappa^2}\right)^{1/4}(2\sqrt{\alpha}\dot{\mc{Z}}+\kappa\dot{\Psi}),
\end{align}
and $\mc{X}=\dot{\mc{Y}}$. Substituting the above equations back into the action \eqref{scal2}, one obtains
\begin{align}\label{scal3}
S^{(2)}_{scalar}&=\f12\int\, d^3k\,dt\,\Bigg[\f{16}{5}(24\beta\kappa^6)^{1/4}a k^2(2 \Psi\dot{\mathcal{Z}}+ \mathcal{Z} \dot\Psi )+\frac{48}{5} \mathit{a}^3 \kappa(\sqrt{\alpha }\dot\Psi 
\dot{\mathcal{Z}}-\kappa \dot\Psi {}^2)-\frac{12}{5} \alpha  \mathit{a}^3 \dot{\mathcal{Z}}{}^2\nonumber\\&-\f{16\kappa}{5\alpha}\sqrt{\f{2\beta}{3}}\f{k^4}{a}(3\kappa^2\Psi^2+4\alpha\mc{Z}^2-4\sqrt{\alpha}\kappa\Psi\mc{Z})+\f{8}{15}ak^2(2\kappa^2\Psi^2-\alpha\mc{Z}^2+10\sqrt{\alpha}\kappa\Psi\mc{Z})\Bigg].
\end{align}
Upon transforming the perturbation variables $\Psi$ and $\mc{Z}$ as
\begin{align}
\mc{N}=\Psi+\frac{2\kappa}{\sqrt{\alpha}}\mc{Z},\qquad
\mc{M}=\Psi-\frac{\sqrt{\alpha}}{2\kappa}\mc{Z},
\end{align}
one can see that the variable $\mc{N}$ becomes non-dynamical. After obtaining the equation of motion for $\mc{N}$ and substituting back to the action \eqref{scal3}, one obtains
\begin{align}
S^{(2)}_{scalar}&=\int\, d^3k\,dt\,\f{8}{3}\kappa^2a^3 \left(9\dot{\mc{M}}{}^2+2\f{k ^2}{a^2} \mc{M}^2\right),
\end{align}
showing that the remaining scalar perturbation suffers from gradient instability. This can be traced back to the fact that the sign of the kinetic term for the scalar field is positive.
\section{Conclusions}
In this paper we have considered the cosmological implications of a theory consists of a scalar field in curved space-time coupled non-minimally to the Maxwell field. The scalar term has a ``John'' self interaction form of the beyond Fab-four theory and the non-minimal coupling between the scalar and the vector field is the interaction between the kinetic term of the scalar field with the energy-momentum tensor of the Maxwell field. In fact the beyond Fab-four Lagrangians do not satisfy recent observational data on the gravitational waves indicating that the speed of the tensor perturbations should be equal to the speed of light \cite{gw}. As a result we have added another self-interaction term to the action which turn it to a subclass of the GLPV theory \cite{glpv} with a speed of tensor mode equal to the speed of light. The theory has two internal symmetries; the translational symmetry associated with the scalar field and the $U(1)$ symmetry associated with the Maxwell field. One should note that the terms appearing in the action \eqref{ac2} can also be found in the Stueckelberg transformation of the beyond generalized Proca theory \cite{beygenproca}. However, our theory is not a special case of the beyond generalized Proca theory since there is no combination of beyond generalized Proca interactions that gives the action \eqref{ac2}. Cosmological consequence of the generalized Proca theory is considered in \cite{cosbeygenproca}.

The theory has two independent exact de Sitter solutions; one is driven by the cosmological constant and is equivalent to the de Sitter solution of the Einstein-Hilbert action. The other is driven by a non-constant, time dependent scalar field. This solution does not need a cosmological constant but the coupling constant $\alpha$ for the canonical kinetic term of the scalar field should be positive.

The dynamical system analysis of the theory shows that the system is one dimensional and has four fixed points. The $\Lambda$-based and GLPV-based solutions coincides with two stable dS fixed points of the theory. There are also two unstable matter dominated fixed points in which the dynamical evolution of the universe can start, and end at the stable $\Lambda$-based dS fixed point at late times.

The cosmological perturbations over these solutions shows that the $\Lambda$-based solution is healthy at linear level for all perturbations provided that the constant $\alpha$ becomes negative. This is in fact satisfactory because in this case all the higher derivative self-interaction of the scalar field vanishes and we left only with a standard kinetic term of the theory. As a result for a healthy scalar perturbations around the $\Lambda$-based solution one should have a correct sign for the scalar's canonical kinetic term. We will then have two branches of dS solutions in this theory. For $\alpha<0$ we have just the $\Lambda$-based solution and for the $\alpha>0$ we have only the GLPV-based solution.

The GLPV-based solution has a healthy tensor perturbations. Also the vector sector, put an lower bound on the values of $\beta$. However, the scalar sector shows a gradient instability which can be traced back to the sign of $\alpha$. In fact, the presence of the GLPV interaction can not compensate the $\alpha$-term in the action and wrong sign of $\alpha$ affect the perturbations at linear level. As a result, one can see that the GLPV-Maxwell system does not have a healthy dS solution without the cosmological constant.
\begin{acknowledgments}
	The author would like to thank the anonymous referee for very useful and important comments.
\end{acknowledgments}

\end{document}